\documentclass[a4paper,fleqn,usenatbib]{mnras}

\usepackage[T1]{fontenc}
\usepackage{ae,aecompl}

\usepackage{graphicx}	
\usepackage{amsmath}	
\usepackage{amssymb}	
\usepackage{multirow}

\def \halpha {H$\alpha$}
\def \kms {{ \rm km\;s$^{-1}$}}

\def \arcsec {$^{''}$}
\def \siiv {Si\,{\sc iv}}

\graphicspath{{figs/}} 

\title[Active region explosive events observed by IRIS and SST/CRISP]{Explosive events in active region observed by IRIS and SST/CRISP}
\author[Z. Huang et al.]{Z. Huang,$^1$ M.~S. Madjarska,$^{2,1,3}$ E.~M. Scullion,$^{4}$ L.-D. Xia,$^{1}$ J. G. Doyle,$^{2}$ T. Ray$^{5}$
\\
$^{1}$Shandong Provincial Key Laboratory of Optical Astronomy and Solar-Terrestrial Environment, Institute of Space Sciences,\\ Shandong University, Weihai, 264209 Shandong, China~[z.huang, xld@sdu.edu.cn]\\
$^{2}$Armagh Observatory and Planetarium, College Hill, Armagh BT61 9DG, UK~[madj, jgd@arm.ac.uk]\\
$^{3}$Max Planck Institute for Solar System Research, Justus-von-Liebig-Weg 3, 37077, G\"ottingen, Germany~[madjarska@mps.mpg.de]\\
$^{4}$Department of Mathematics \& Information Sciences, Northumbria University, Newcastle Upon Tyne, NE1 8ST, UK~[scullie@tcd.ie]\\
$^{5}$Dublin Institute for Advanced Studies, 31 Fitzwilliam Place, Dublin 2, Ireland~[tr@cp.dias.ie] }

\date{Accepted XXX. Received YYY; in original form ZZZ}

\pubyear{2016}

\begin{document}
\label{firstpage}
\pagerange{\pageref{firstpage}--\pageref{lastpage}}
\maketitle

\begin{abstract}
Transition-region explosive events (EEs) are characterized by non-Gaussian line profiles  with enhanced wings at Doppler velocities of 50--150\,\kms. They are believed to be the signature of solar phenomena that are one of the main contributors to coronal heating. The aim of this study is to investigate  the link of EEs to dynamic phenomena in the transition region and chromosphere in an active region.  We analyze observations simultaneously taken by the Interface Region Imaging Spectrograph (IRIS) in the \siiv\ 1394\,\AA\ line and the slit-jaw (SJ) 1400\,\AA\ images, and the Swedish 1-m Solar Telescope (SST) in the  \halpha\ line.  In total 24 events were found. They are associated with small-scale loop brightenings in SJ~1400\,\AA\ images. Only four events show a counterpart in the  \halpha$-35$\,\kms\ and \halpha$+35$\,\kms\ images. Two of them  represent brightenings in the conjunction region of several loops that are also related to a bright region (granular lane) in the \halpha$-35$\,\kms\ and \halpha$+35$\,\kms\ images.  Sixteen  are general loop brightenings that do not show any discernible response in the \halpha\ images. Six EEs appear as propagating loop brightenings, from which two are associated with dark jet-like features clearly seen in the \halpha\,$-35$\,\kms\ images. We found that chromospheric events with jet-like appearance seen in the wings of the \halpha\ line can trigger EEs in the transition region and in this case the IRIS \siiv\ 1394\,\AA\  line profiles are seeded with  absorption components resulting from Fe\,{\sc ii} and Ni\,{\sc ii}. Our study indicates that EEs occurring in active regions have mostly upper-chromosphere/transition-region origin. We suggest that  magnetic reconnection resulting from the braidings of small-scale transition region loops is  one of the possible mechanisms of energy release that are responsible for the EEs reported in this paper.
\end{abstract}

\begin{keywords}
Sun: activity - Sun: chromosphere - Sun: transition region - techniques: spectroscopic - methods: observational
\end{keywords}

\section{Introduction}
\label{sect_intro}
Explosive events (EEs) are small-scale transients (duration~$<$600\,s, size~$<$5\arcsec) observed in the solar transition region, and describe non-Gaussian line profiles  with enhanced wings at Doppler velocities of 50--150~\kms\,\citep{1983ApJ...272..329B,1989SoPh..123...41D}.
The physical nature of EEs is under investigation for more than two decades. EEs were firstly suggested to be the spectral footprint of bi-directional jets caused by magnetic reconnection\,\citep{1991JGR....96.9399D,1997Natur.386..811I}. Later, chromospheric upflow events \citep{1998ApJ...504L.123C}, siphon flows in small-scale loops \citep{2004A&A...427.1065T}, surges\,\citep{2009ApJ...701..253M} and transient brightenings and X-ray jets \citep{2012A&A...545A..67M} have been associated with EEs.

\par
\citet{1991ApJ...370..775P} reported that  explosive events occur in the solar magnetic network lanes. This has been later confirmed by many further studies \citep[e.g.][etc.]{2003A&A...403..731M,2004A&A...419.1141N,2004A&A...427.1065T,2008ApJ...687.1398M}. \citet{1998ApJ...497L.109C} established that the majority of explosive events are associated with the cancellation of photospheric magnetic flux, which was  recently confirmed by \citet{2014ApJ...797...88H} and \citet{2015ApJ...809...82G}. EEs were modelled by \citet{1999SoPh..185..127I}, \citet{2002A&A...383..697R}, \citet{2001A&A...380..719R}, \citet{2001A&A...375..228R}, \citet{2001A&A...370..298R}, and \citet{2015arXiv150908837I} in two-dimensional  numerical simulations as the product of magnetic reconnection.

\par
Non-Gaussian line profiles in the solar transition region are intensively investigated since the first flight of the Naval Research Laboratory (NRL)  High Resolution Telescope and Spectrograph (HRTS)  in 1975. \citet{1989SoPh..123...41D} reported on the various shapes of C~{\sc iv} line profiles (see their Figures~2, 3, 6, 8, 9, 11, 13, and 15) that were named as ``explosive events''.  Later \citet{1991JGR....96.9399D} suggested that EE line profiles are the spectroscopic signature of magnetic reconnection. Recently, \citet{2014Sci...346C.315P} reported similar Si~{\sc iv} line profiles observed by \textit{IRIS} with superimposed absorption lines from singly ionised ion, neutral atom and/or molecular lines  \citep{2014A&A...569L...7S}  in active region dot-like events that the authors called ``hot explosions''.  The electron density of the hot explosion was estimated to exceed $10^{13}$\,cm$^{-3}$. \citet{2014Sci...346C.315P}  suggested that these line profiles are the products of small-scale magnetic reconnection occurring in  the photosphere while  \citet{2015ApJ...808..116J} put forward the idea that they are rather the product of Alfv\'enic turbulence originating in the chromosphere or above.

\par
The Interface Region Imaging Spectrograph\,\citep[IRIS,][]{2014SoPh..tmp...25D} launched in 2013 obtains observations of the solar transition region at  unprecedented spatial and spectral resolution providing great opportunities to investigate the physical processes that generate EEs. \citet{2014ApJ...797...88H} studied an EE that occurred at the boundary of the quiet Sun and a coronal hole observed with {\it IRIS}, the Atmospheric Imaging Assembly {\it (AIA)} and the Helioseismic and Magnetic Imager {\it (HMI)}. The authors found that the EE is associated with  a complex loop system seen in the AIA 171\,\AA\ passband.  The magnetic cancellation rate during the event was of $5\times10^{14}$\,Mx\,s$^{-1}$. The EE reached a temperature of at least $2.3~\times~10^5$\,K. \citet{2014ApJ...797...88H} suggested that this EE is caused by magnetic reconnection within the complex loop system. A recent study by \citet{2015ApJ...810...46H} found  EEs  in a footpoint of a cool transition region loop system and also at the  footpoint junction of two loop systems where cancelling opposite magnetic polarities were present. \citet{2015arXiv150908837I} reported IRIS observations of 15 explosive events occurred in active region. They analyzed in detail the core and wing emission of the explosive event spectra, and found that core and wing emission are spatially coincident, and do not move significantly during the typical event duration. The numerical experiment shows that multiple magnetic islands and acceleration sites characterising the plasmoid instability in fast magnetic reconnection can reproduce the observed profiles. \citet{2015ApJ...809...82G} found short-period variability (30 s and 60--90 s) within EE bursts in IRIS observations.

\par
In the present work we aim to investigate the link of EEs  observed in the transition region by IRIS in a Si~{\sc iv} line to  chromospheric phenomena registered in the wings of H$\alpha$ spectral imaging data. After identifying the EEs in the \siiv\ 1394~\AA\ line, the IRIS slit-jaw images at 1400\,\AA\ passband and \halpha\ images taken by the CRisp Imaging SpectroPolarimeter \,\citep[CRISP,][]{2008ApJ...689L..69S} installed at the Swedish 1-m Solar Telescope\,\citep[SST,][]{2003SPIE.4853..341S} in La Palma are used to identify the \halpha\ wing counterparts of the EEs. 

\par
In the following, we describe the observations and the data analysis in Section\,\ref{sect_obs}.  The results and discussion are presented in Section\,\ref{sect_res}. We give the conclusions in Section\,\ref{sect_concl}.

\begin{figure}
\includegraphics[width=0.5\textwidth,clip,trim=0.5cm 1cm .5cm 2cm]{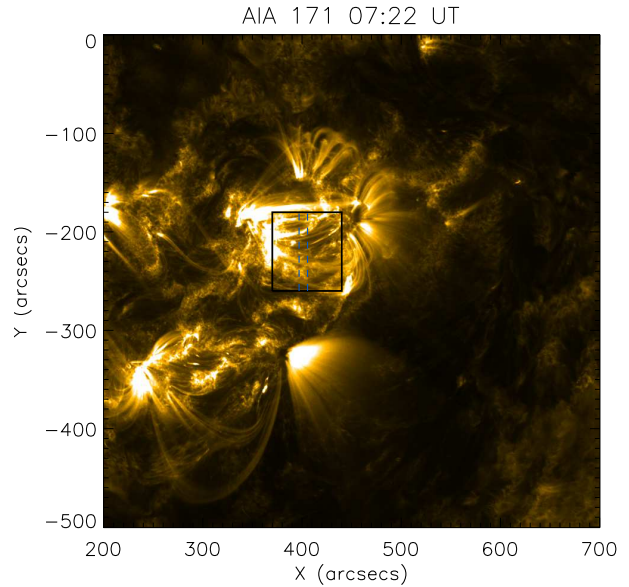}
\caption{Partial view of the solar disk observed in the AIA 171\,\AA\ passband. The analyzed IRIS and CRISP field-of-views are denoted by the black box. The region between the dashed lines (in blue) is the place where the IRIS spectra were obtained.}
\label{fig_fov}
\end{figure}

\begin{figure}
\includegraphics[width=0.5\textwidth,clip,trim=0.1cm 1.8cm 0cm 0.5cm]{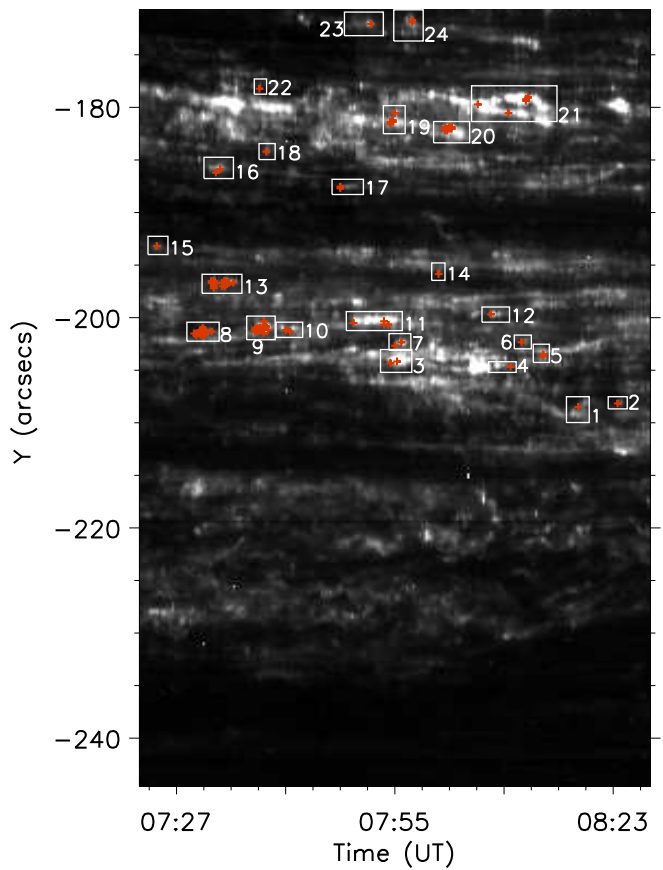}
\caption{\siiv\ 1394\,\AA\ radiance image of the region observed by the IRIS in the quasi ``sit-and-stare'' mode. The plus symbols are the locations of EE line profiles detected by the automatic procedure. The red symbols present those detected by step 3 and the green ones are from step 5. The selected events (24 in total) are denoted by boxes with numbers.}
\label{fig_ee_id}
\end{figure}

\begin{figure*}
\includegraphics[width=17cm,clip,trim=0cm 0.1cm 0cm 0.1cm]{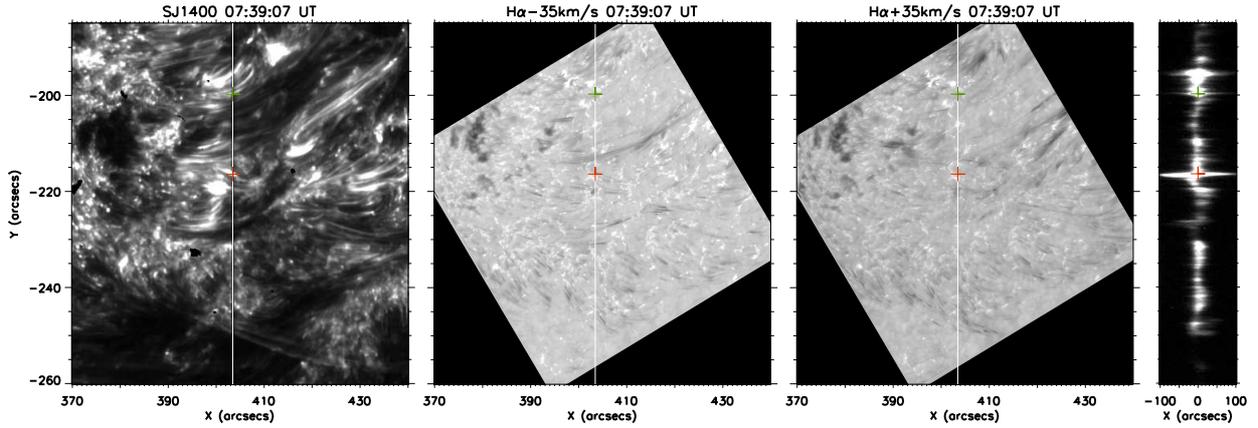}
\caption{ Locations of the EE line profiles (red plus signs: identified by step 3 of the automatic method, green plus signs: identified by step 5 of the automatic method) superimposed on the IRIS SJ~1400\,\AA\ (the first panel), the \halpha\ $-35$\,\kms\ (the second panel) and \halpha\ $+35$\,\kms\ (the third panel) images. White solid lines denote the slit of the IRIS spectrograph. The slit image is shown in the fourth panel, where the spectral range is from $-100$\,\kms\ to 100\,\kms, in order to clearly present the wing enhancement. (An animation of this figure is available online).}
\label{fig_ee_on_map}
\end{figure*}

\begin{figure*}
\includegraphics[width=17cm,clip,trim=1.8cm 2.2cm 6cm 0cm]{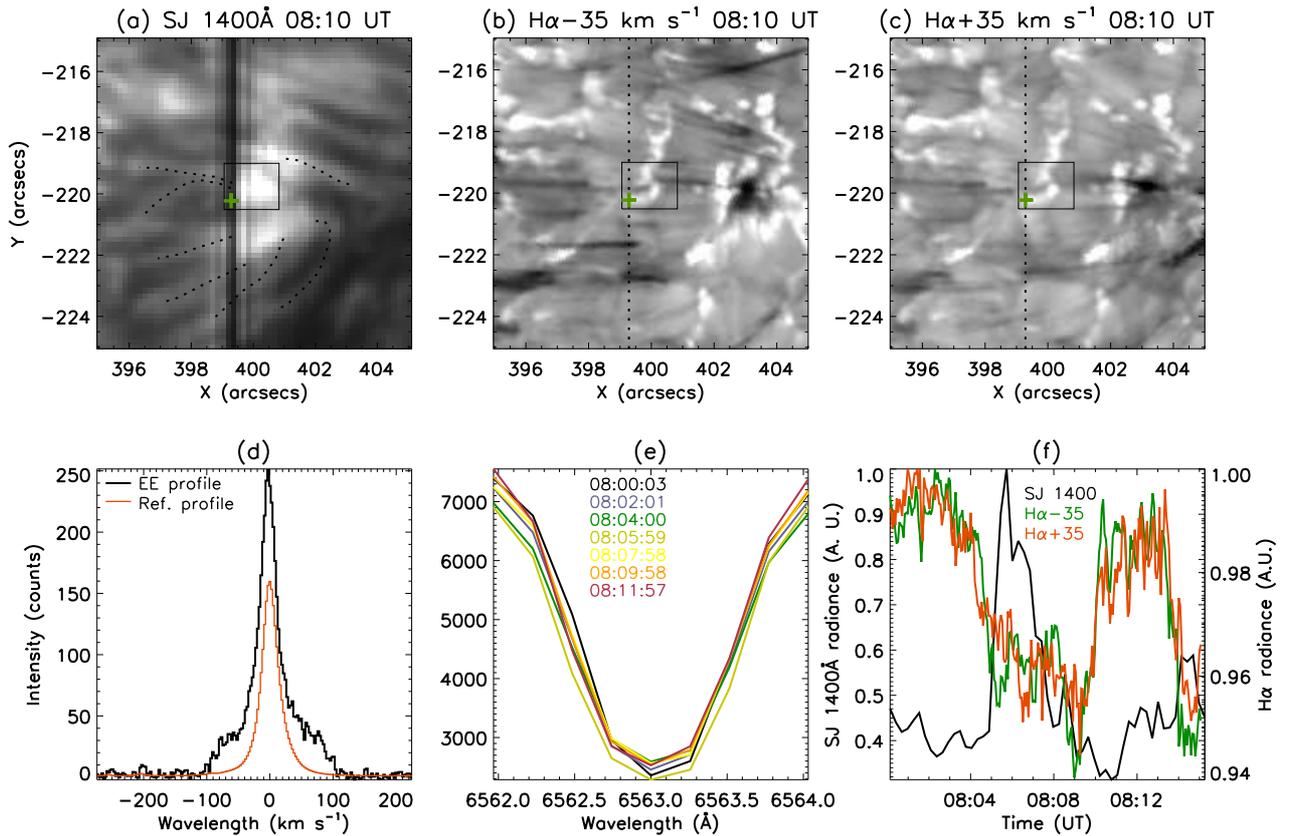}
\caption{\siiv\ 1394\,\AA\ EE spectra found in a conjunction region of the footpoints of a few loop systems (EVENT 4):
The top row shows the conjunction region viewed in  the IRIS SJ~1400\,\AA\ (a), CRISP \halpha$-35$\,\kms\ (b) and  \halpha$+35$\,\kms\ (c). The dotted lines (in black) in panel (a) outline some of the loops connecting in the conjunction region, and the dark vertical bar is the location where the light is blocked by the IRIS spectrometer slit, which is also over-plotted as dotted line in panels (b) and (c). The box (in black) denotes the region from which the lightcurves shown in panels (f) are generated. The plus sign (in green) denotes the location of the  EE \siiv\ spectrum shown in panel (d).
Bottom row: (d) \siiv\ 1394\,\AA\ line profile (in black) taken from the location marked by a plus sign in the top row together with an average spectrum (in red) taken from the full field of view shown in Fig.\,\ref{fig_ee_id}; (e) \halpha\ line profile from the same location where the EE \siiv\ spectrum emitted. The observed time of each profile is labeled; (f) the lightcurves of the boxed region in the SJ 1400\,\AA\ (black), \halpha$-35$\,\kms\ (green), \halpha$+35$\,\kms\ (red).}
\label{fig_no4}
\end{figure*}

\begin{figure*}
\includegraphics[width=17cm,clip,trim=1.8cm 3.8cm 0.5cm 0cm]{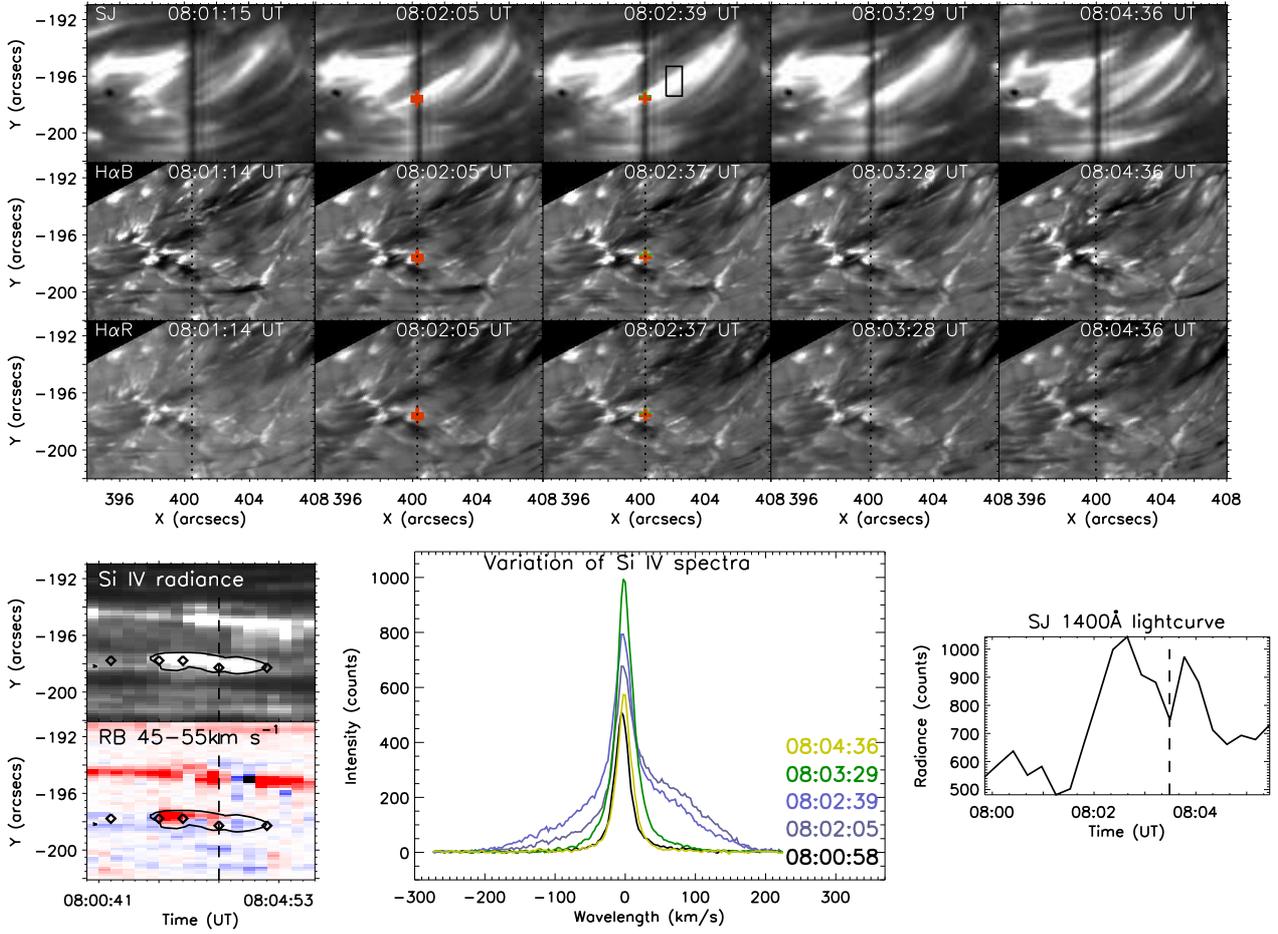}
\caption{Si~{\sc iv} 1394\,\AA\ EE spectra found in a loop brightening (EVENT 20). From top to bottom, first three rows show the evolution of the loop region seen in the IRIS SJ~1400\,\AA, \halpha$-35$\,\kms\ (\halpha B) and \halpha$+35$\,\kms\ (\halpha R), respectively. The dotted lines on the \halpha\ images  show the locations of the spectrometer slit. The plus signs (in red and green) are the locations of the identified EE spectra. The black box on the SJ image at 08:03:29\,UT denotes the region, from which the lightcurve (bottom row of this figure) is obtained. Bottom row:  The left panel displays the \siiv\ radiance image (top) and RB asymmetry at 45--55~\kms\ (bottom) of the region, in which the event is outlined by a contour line (black solid line). The spectra  shown in the middle panel are taken from  the pixels marked with diamond symbols in the left panel. The dashed line denotes the time (08:03:29\,UT) when the \siiv\ spectra turns to single-Gaussian. The middle panel displays the variation of the \siiv\ spectra taken from the pixels shown in the left panel (time is denoted by colors). The right panel shows a SJ 1400\,\AA\ lightcurve taken from the box region marked in the SJ 1400\,\AA\ image, where the dashed line marks the time at 08:03:29~UT.}
\label{fig_no20}
\end{figure*}

\begin{figure*}
\includegraphics[width=17cm,clip,trim=0.5cm 4.5cm 0.5cm 0cm]{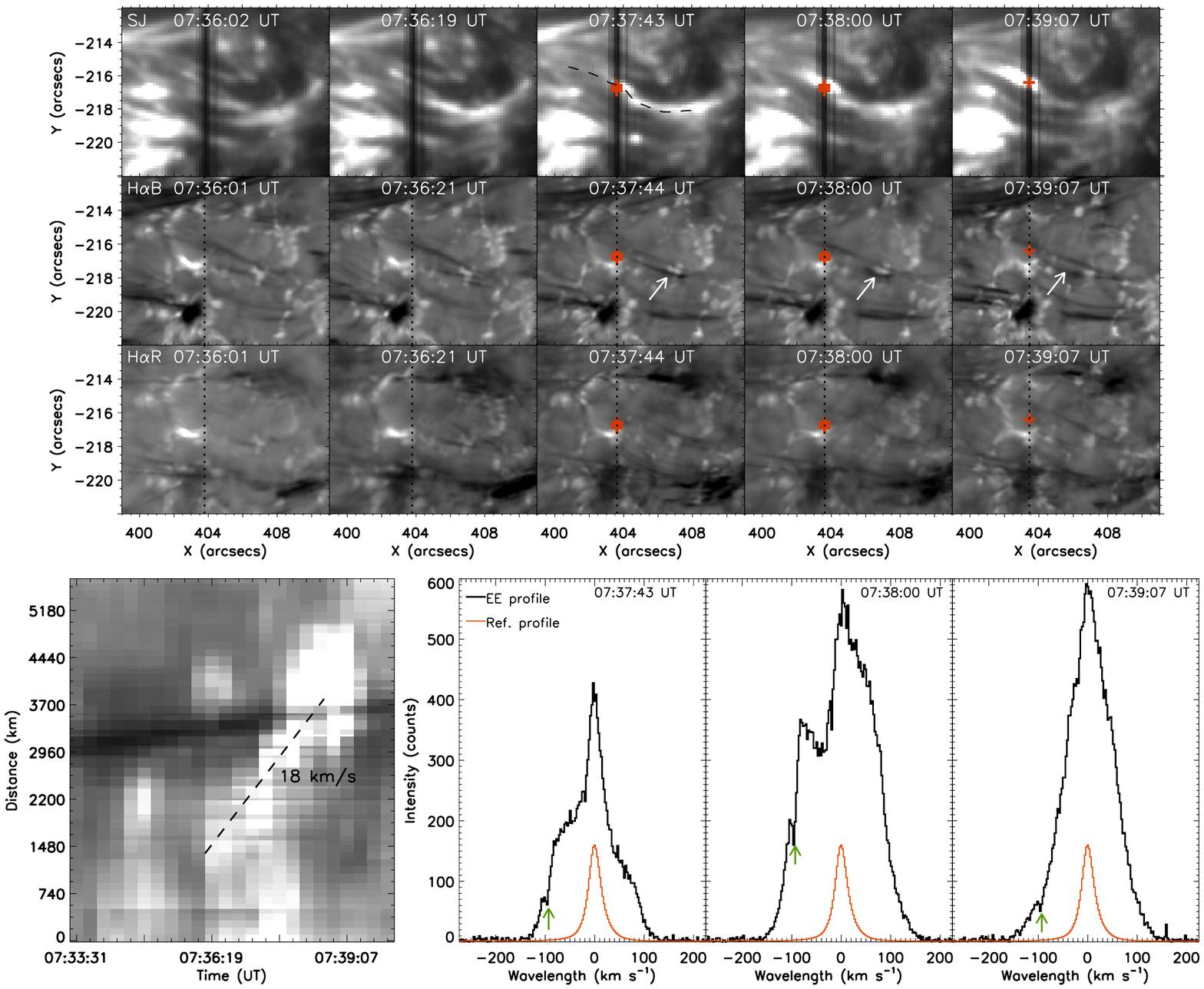}
\caption{EE spectra found in a propagating loop brightening. From top to bottom, first three rows show evolution of the region seen in the IRIS SJ~1400\,\AA, \halpha$-35$\,\kms\ (\halpha B) and \halpha$+35$\,\kms\ (\halpha R) respectively. The dotted lines on the \halpha\ images denote the location of the spectrometer slit. The plus signs (in red) are the locations of the identified EE spectra. A bright structure is found to move along a sigmoid path marked with a dashed line on the SJ~1400\,\AA\ image at 07:37:43\,UT. A time-slice image along this path is given in the bottom panel of this figure.
Bottom row: The left panel displays the time-slice plot from the cut denoted on the SJ~1400\,\AA\ image at 07:37:43\,UT. The right panels show EE spectra at different times obtained from the locations marked by plus symbols on the SJ images. The arrows denote the absorption components in the \siiv\ spectra, and the reference profile is shown in red.}
\label{fig_no9}
\end{figure*}

\section{Observations and data analysis} 
\label{sect_obs}
The data used in the present study were taken by IRIS and CRISP on 2014 June 10 from 07:22\,UT to 08:28\,UT (the CRISP observations end at 08:23\,UT), targeting active region NOAA 12085 (see Fig.\,\ref{fig_fov}). 

\par
The IRIS spectral data were obtained in a sit-and-stare mode with a 17\,s cadence and a 15\,s exposure time. The size of the IRIS slit is 0.35\arcsec$\times$129\arcsec  (only part of the  field-of-view  of the data are analyzed because the CRISP field-of-view is smaller). Seven spectral windows were transferred to the ground, and in this study we only use the \siiv\,1394\,\AA\ line (T$_{max}\sim$ 63\,000\,K) to identify EEs in the transition region. The IRIS slit-jaw (SJ) images were recorded in the 1400\,\AA\ passband with a cadence of 17\,s and a 119\arcsec$\times$129\arcsec field-of-view. The pixel size of the IRIS slit-jaw images is 0.17\arcsec$\times$0.17\arcsec. The IRIS data were downloaded as level 2 products that have been reduced by the instrument team and no further reduction procedure is required for the purpose of this study. Compensation for the solar rotation was not applied to this dataset, therefore the IRIS slit actually scanned 1\arcsec\ from the solar surface in 447~s, i.e. $\sim$26 exposures.

\par
CRISP collected the imaging spectroscopy in \halpha\ (line center at 6563\,\AA), scanning the line profile from 6561.97\,\AA\ to 6564.03\,\AA\ in 9 steps at a 0.26\,\AA\ step size. The CRISP field-of-view (FOV) is 60\arcsec$\times$60\arcsec with a pixel scale of 0.0585\arcsec and  a cadence of  about 4\,s. The data were reduced with the CRISPRED package\,\citep{2015A&A...573A..40D}.

\par
A comparison between the IRIS SJ 1400\,\AA\ image and the \halpha\ images taken at $\pm$35~\kms\ blue and red wings are used to align the data taken with the two instruments. This alignment also gives the image tilt and the pixel sampling scale of CRISP. 

\par
In order to identify explosive event profiles in the observed field-of-view, we tested several methods that can be applied for large rasters or sit-and-stare data (i.e. numerous line profiles). We used a triple Gaussian fit (the procedure developed by P. Young: \url{http://pyoung.org/quick_guides/iris_auto_fit.html}) as well as mapping at the different Doppler shifts in the blue and red wings. Both methods identified numerous pixels with strong line shifts. The visual inspection, however, showed that a lot of these pixels are actually mis-identifications because of instrumental noise or spiked pixels caused by cosmic rays or other non-solar artefacts. To eliminate these pixels we developed an automatic method to select the true \siiv\,1394\,\AA\ spectra with enhanced wings at both blue and red ends (Doppler shifts larger than 50\,\kms). For details of the identification method, please refer to  the Appendix \ref{sect_eeid}. All the identifications are then visually checked to make sure they fulfil the characteristics of explosive event spectra\,\citep[see][]{1989SoPh..123...41D}.

\par
Once the non-Gaussian profiles are identified, their locations are superimposed on  the \siiv\ radiance image of the sit-and-stare observations (Fig.\,\ref{fig_ee_id}). Those spectra that are temporally and spatially clustering together are grouped as one event (see Fig.\,\ref{fig_ee_id}). In some bright patches, only a few pixels are identified as the rest do not meet our criteria. Nevertheless, the whole bright patch is defined as one event. In total, 24 EEs are selected for further investigation (see Fig.\,\ref{fig_ee_id}). The locations of the EEs were then superimposed on the  SJ images and  on the aligned CRISP \halpha\ images.

\section{Results and discussion}
\label{sect_res}
In Fig.\,\ref{fig_ee_id}  we display the \siiv\ radiance image of the sit-and-stare observations with the locations of the detected EE line profiles. The EE spectra are mostly found in bright patches and their boundaries as seen in the \siiv\ radiance map.

\par
The locations of the detected EE line profiles on the IRIS SJ 1400\,\AA\ images and the CRISP images in the  \halpha $-35$\,\kms\ and \halpha $+35$~\kms\ wings are shown in Fig.\,\ref{fig_ee_on_map} and the online animation. First, we identified the features associated with the EE spectra in the SJ 1400\,\AA\ images, and then mapped them on the \halpha\ images. The visual analysis of these  features revealed that the EE spectra are always associated with brightenings observed in the SJ 1400\,\AA\ images. The brightenings are found in loops and  at the footpoint conjunction of several loops (see Table\,\ref{tab_eeresp}). Most (16 out of 24) of the bright patches are associated with general loop brightenings, 2 out of 24 are found in the footpoint conjunction region of several loops, and 6 out of 24 are related to propagating brightenings along small-scale loops.  EEs are predominantly found in the network regions (see Section\,\ref{sect_intro}) where loops are rooted. Because of the small field-of-view of our data, just two EEs are found in loop footpoints and are associated with compact brightenings in the SJ images. Our observations clearly show that EEs can also be associated with  brightenings along loops.

\par
The location of each EE were then investigated in the \halpha$\pm$35\kms\ images.  If an \halpha\ absorption or emission phenomenon was detected  in a given location and shows  similar structure and evolution as  seen in the SJ 1400\,\AA\ images, we assume that the EE has an  \halpha\ wing counterpart. Please note that no features were selected in the \halpha\ line center as the region viewed in the wavelength is crowded with both filamentary  bright and dark features. However, we cannot exclude the possibility that   a feature is present at this wavelength but is not distinguishable from the abundance of features seen there. Thus, the present study focussed  on the dynamic phenomena that are only seen in the \halpha\ line wings.  

 In Table\,\ref{tab_eeresp} we give the grouping  of the 24 events (Fig.\,\ref{fig_ee_id}) according to their appearance  in the SJ\,1400\,\AA\ and the \halpha$\pm$35\kms\ images. Four events have a response in the \halpha$\pm$35\,\kms\ images.   In the \halpha$\pm$35\,\kms\ images, events No.~3 \& 4  are associated with a bright granular lane where the loops are possibly rooted, events No. 8 \& 9  appear  to link to dark jet-like features, and the rest (18 of 22) do not show any detectable response  in the  \halpha\ wings  (see Table\,\ref{tab_eeresp}). In the following sections, we give examples of the different groups of features.

\begin{table*}
\caption{EE corresponding events in the SJ~1400\,\AA\ and \halpha$\pm$35\,\kms\ images.}
\label{tab_eeresp}
\begin{tabular}{|c|c|c|c|}
\hline
Wavelength (passbands)&Features in the passband&Indices of events&Total\\
\hline
\multirow{3}{*}{SJ 1400\,\AA}&Conjunction region of a few loops&3, 4&2/24\\
\cline{2-4}
&General loop brightenings&1, 2, 7, 12--24&16/24\\
\cline{2-4}
&Propagating loop brightenings&5, 6, 8--11&6/24\\
\hline
\multirow{4}{3cm}{\halpha$-$35\,\kms\ \& \halpha$+35$\,\kms}&Bright granular lane&3, 4&2/22\\
\cline{2-4}
&Dark jet-like feature&8, 9&2/22\\
\cline{2-4}
&None&All others& 18/22\\
\hline
\multicolumn{4}{|c|}{Note: events No. 23 \& 24 are not in the CRISP FOV.}\\
\hline
\end{tabular}
\end{table*}

\subsection{ EE spectra at conjunction areas}
\label{sect_no3}
Events No.~3 and 4 are found in the footpoint conjunction area of loops. The response of the two events can be seen in the \halpha\ blue ($-35$\,\kms) and red ($+35$\,\kms) wing images. Here, we present event No.~4  in detail. Several loops root into a bright region showing dynamic evolution  seen in IRIS SJ 1400\,\AA\ (Fig.\,\ref{fig_no4}a and online animation). An EE spectrum (Fig.\,\ref{fig_no4}d) is identified at the east edge of the bright region. The spectra in the rest of the bright region are not identified as EEs by our automatic procedure, although most of them show enhanced line wings,  but their Doppler shifts  and intensities are lower than our criteria. Small-scale brightenings along the loops can be seen (see the online animation) in only one or two frames (17\,s cadence), which, unfortunately, does not allow a reliable propagation speed estimation.

\par
The region seen in the \halpha\ images is shown in Figs.\,\ref{fig_no4}b--c. A bright feature,  a granular lane, is clearly seen in the \halpha$-35$\,\kms\ and \halpha$+35$\,\kms\ images.   The \halpha\ line profiles obtained from the same location of the EE do not show any significant change during the EE scanned by the IRIS slit (see Fig.\,\ref{fig_no4}e). The lightcurves of the whole bright patches at \halpha$-35$\,\kms\ and $+35$\,\kms\ present a small ($\sim$6\%) intensity decrease while the SJ~1400\,\AA\ peaks around 08:06\,UT (Fig.\,\ref{fig_no4}f). Please note that the rapid decrease in the SJ~1400\,\AA\ lightcurve after 08:06\,UT is largely instrumental and is caused by the spectral slit blocking the light. A direct comparison of the lightcurves in Fig.\,\ref{fig_no4}f should be, therefore, made with caution. When cross-checking with the animation, we found that the decrease in the~\halpha\ lightcurves is caused by a dark feature moving across the selected field-of-view. This dark feature, however, is not related to the EE since it appears to initiate away from the bright granular lane region. To conclude, the EE is associated with a bright granular lane in the chromosphere, but the EE activity does not correspond to any  dynamic phenomenon in the \halpha\ wing observations.

\par
While investigating the \siiv\,1394\,\AA\ spectra in this conjunction area, we found that none of the  spectra shows similar properties as those presented in \citet{2014Sci...346C.315P}. If magnetic reconnection is responsible  for the observed spectra in this area, it appears to take place in the higher atmosphere, i.e. the upper chromosphere/transition region that cannot give rise to absorption components \citep[Fe\,{\sc ii}, Ni\,{\sc ii} and/or molecular lines,][]{2014A&A...569L...7S} in the \siiv\ spectra.

\subsection{EE spectra found in loop brightenings}
\label{sect_no20}
Most (67\%) of the EE spectra are  associated with general loop brightenings seen in the  SJ~1400\,\AA\ images. These brightenings are seen to extend along the loop length, and the intensity at different locations flares up simultaneously. The apparent motion of the events cannot be measured because  of the relatively low cadence of the observations. None of these events shows a counterpart in the \halpha\ wing images (see Table\,\ref{tab_eeresp}). In this section, we describe one of these events (No.\,20) in detail.

\par
Fig.\,\ref{fig_no20} displays the evolution of event No. 20. The EE spectra are detected in the observations from 08:02\,UT to 08:03\,UT, when a loop brightening appears at the EE location in the SJ~1400\,\AA\ images. The length of the brightening is 8\arcsec--10\arcsec, and the emission along the loop   seems to become enhanced simultaneously (at 17\,s cadence). The brightened loops show no response in the \halpha$\pm35$\,\kms\ images. 
\par
The EE spectra in this event show enhancements in both the blue and the red wings at similar Doppler shifts (see the examples at 08:02:05\,UT and 08:03:29\,UT shown in Fig.\,\ref{fig_no20}), but the emission enhancement in the red wing is much stronger. Another interesting feature of this event is that the \siiv\,1394\,\AA\ spectra turn to single-Gaussian profiles after 08:03:29\,UT, even though the loop is still bright and the emission in the \siiv\ is very strong (see the \siiv\ radiance map and the spectra at 08:03:29\,UT and 08:04:36\,UT in Fig.\,\ref{fig_no20}). The lightcurve obtained from a region of the bright loop in the SJ~1400\,\AA\ images (sampled at a place that is not affected by the spectrometer slit) is also shown in Fig.\,\ref{fig_no20}. It shows a dynamic evolution in the loops, but the EE is observed only during a short period of this evolution. This indicates that not all loop brightenings emit EE spectra.

\par
The brightenings witnessed by EEs and their association with small-scale loops suggest energy release most probably produced by magnetic reconnection that results from small-scale loop braidings driven by the continuous footpoint motions \,\citep{1983ApJ...264..642P, 1988ApJ...330..474P}. While braidings are not everywhere along the loops, it explains why some brightenings in the same loops do not produce EE spectra. Multiple threads of loops can be seen in the region (see the top panel of Fig.\,\ref{fig_no20}), which implies a complex loop system providing favorable conditions for braiding. Series of studies  by \citet{2010A&A...516A...5W}, \citet{2011A&A...525A..57P} and~\citet{ 2015ApJ...805...47P} (also the references therein) have investigated through 3D MHD experiments the magnetic field braiding  and the magnetic reconnection associated with it. However, none of these experiments so far has provided observables that can be directly compared with spectral (line profiles) or imaging data. 

\par
 Transition region \siiv\ line profiles similar to the EE profiles shown here were also reported by \citet{2014Sci...346D.315D} and were associated with what appears to be a twisted feature identified by one side of the feature being more red-shifted and the other more blue-shifted, resulting in tilted emission along the spectral slit. While carefully checking the spectral profiles in the brightening loops studied here, we also found some cases with a tilted appearance in the wavelength--space plots.  Such tilted appearance can be demonstrated by a RB asymmetry map (see the description of the technique in e.g.,  \citet{2009ApJ...701L...1D} and \citet{2011ApJ...738...18T}) at 45--55~\kms\ of this region (see bottom row of Fig.\,\ref{fig_no20}). In the RB asymmetry plot, such a tilt is clearly present between 08:02:39\,UT and 08:03:29\,UT, when the top part of the brightening loops is redshift-dominated and the bottom part is blueshift-dominated. However, since such a feature is not perpetual in the event and none of the events showing such tilt has been associated with a feature seen in \halpha, we cannot speculate on whether the tilt is caused by a twist  as found in \citet{2014Sci...346D.315D}. Nevertheless, the interpretation given by \citet{2014Sci...346D.315D} is worth a further investigation when other similar IRIS and \halpha\ data become available.

\par
Recently, \citet{2014PASJ...66S...7B,2015A&A...580A..72B} reported on a 3D MHD model investigation of the magnetic field-line braiding mechanism as a major source for coronal heating in ARs. With a large-scale horizontal velocity field derived from time series of Hinode/SOT magnetograms and a smaller-scale one added artificially based on granular cell motions, the authors report that the large-scale magnetic field concentration and small-scale  granular motions introduce enough stress at the footpoints of the magnetic field lines. This leads to a continuous dissipation of magnetic energy on small spacial scales (down to 230~km).
A similar approach was used by many other authors\,\citep[and the references therein]{2005ApJ...618.1020G}. Although all these studies refer to large-scale coronal loops, the present observations provide direct evidence that this mechanism might also be  at work in the small-scale cool loops in the transition region above active regions. 

\subsection{EE spectra found in propagating loop brightenings}
\label{sect_jetee}
Of all 24 events, six are found to be related to propagating loop brightenings seen in the SJ~1400\,\AA\ images (see Table\,\ref{tab_eeresp}).  In these cases the apparent velocities of bright plasma blobs can be measured. From these six events, only two  (No. 8\&9) appear to show a counterpart in the \halpha\ blue and red wing images where dark jet-like features are present. 

\par 
In Fig.\,\ref{fig_no9} we show the evolution of the region  where No.~9 event is seen. The EE spectra are associated with a bright feature moving along an S-shaped flux tube seen in the SJ~1400\,\AA\ (see the SJ image at 07:37:43~UT in Fig.\,\ref{fig_no9}). The apparent speed determined from the time-slice plot is estimated at 18\,\kms.

\par
In the \halpha\ blue wing images, dark jet-like features are visible (denoted by arrows in Fig.\,\ref{fig_no9}). These features are very faint in the \halpha\ red-wing images, suggesting that they are mostly upward moving  (blue-shifted) chromospheric plasmas. The dark feature has actually faded away when it reaches the location of the EE. We speculate that the cold plasma captured in the \halpha\ blue wing images may have been heated by the release of energy, e.g. from magnetic reconnection,  and the EE spectra are the indicator of this energy release. The \siiv\ spectra in this event show clear absorption components (see examples given in Fig.\,\ref{fig_no9}), which are produced by the Fe\,{\sc ii} and Ni\,{\sc ii} lines \citep[see,][]{2014A&A...569L...7S,2014Sci...346C.315P}. This absorption components might result from the ejection of cold plasma that is observed as dark jet-like features in the \halpha\ blue wing images. 

\par
Loop brightenings observed in this study have similar  morphology (elongated loop-like brightenings) as the flaring arch filaments  (FAF) observed by \citet{2015ApJ...812...11V}. The \siiv\ 1394\,\AA\ spectra of the two FAFs  reported in \citet{2015ApJ...812...11V} have similar profiles as the ones shown here (Fig.\,\ref{fig_no9}). However, the FAFs show brightenings in the line core, as well as in the blue and red wings of \halpha.  In contrast, the event 9 shown here appears as a dark feature, i.e. seen in absorption in the  \halpha\ wings. The FAFs were suggested to be the signature of magnetic reconnection.  As absorption blends are observed in the \siiv\ 1394\,\AA\ spectra, \citet{2015ApJ...812...11V} suggest two possibilities: either ``the FAFs occur fairly low in the solar atmosphere'' or  ``cool photospheric gas was kicked to the higher atmosphere''. They also pointed out that the second scenario is unlikely because the lack of large Doppler shifts in the blending absorption lines. However, the Doppler shift measurement highly depends on the projection angle, therefore, this scenario is still plausible.  FAFs are phenomena that have not been studied yet in great detail and further investigations are required to explain their nature. Alternatively, the phenomena observed here might also be explained by magnetic reconnection in braiding loops since many intersections between loop threads can be clearly seen (Fig.\,\ref{fig_no9}, top panel, 07:36:02\,UT and 07:36:19\,UT). We can speculate that the propagating brightenings at a speed of 18\,\kms\ can be the result of magnetic reconnection in a chain of braiding knots along the loops occurring one after the other.

\section{Conclusions}
\label{sect_concl}
In the present study, we investigate  transition region explosive events and their possible counterpart in \halpha\ wing images taken at $\pm$35\kms\ by analysing spectral imaging co-observations taken by IRIS and CRISP/SST. We identified 103 non-Gaussian spectra that were then grouped in 24 events. The evolution of these events was analyzed in the IRIS SJ 1400\,\AA\ and CRISP \halpha\ co-observations.

\par
We found that all EE spectra are related to loop brightenings seen in SJ 1400\,\AA\ images. From  24 events, 2 are associated with brightenings in the footpoint  conjunction region of several loops, 16 are found in  general loop brightenings and 6 are related to propagating loop brightenings. Although all the identified EE spectra are related to loop brightenings, not all loop brightenings  seen in the SJ\,1400\,\AA\ images emit EE spectra.

\par
Only four events are found to have a counterpart in the \halpha\ wings.  In the \halpha$-35$\,\kms\ and \halpha$+35$\,\kms\ images, two events are found in the  loop conjunction region  that appear as bright granular lane and two propagating loop brightenings in the SJ~1400\,\AA\ appear to correspond to dark jet-like features. The EE spectra found in general loop brightenings seen in SJ 1400\,\AA\ do not show any discernible response in the \halpha\ wing images. 
Absorption components from Fe\,{\sc ii} and Ni\,{\sc ii} in \siiv\ spectra are found in the events associated with propagating loop brightenings in the SJ~1400\,\AA\ images and dark jet-like features in \halpha\,$-35$\,\kms\ images. The absorption components seeded in the \siiv\ spectra might be the indication of the ejection of cold plasma observed as dark jet-like features in the \halpha\ blue wing images.  \citet{1998ApJ...504L.123C} reported that upflow chromospheric events (observed with the \halpha$-$0.5~\AA\ in the BBSO spectrograph) in the quiet Sun are associated with Si~{\sc iv} 1402.8~\AA\ explosive events (observed with the Solar Ultraviolet Measurements of Emitted Radiation (SUMER) spectrograph). The chromospheric upflow events were suggested by the author to be the manifestation of cool plasma material flowing into magnetically diffusive regions, while explosive events were believed to be hot plasma material flowing out of the same regions. The second part of our investigation in the quiet-Sun region\,\citep{madj2016} will shed more light on this issue.

\par
Our study indicates that explosive events in active regions are associated with transition-region loop brightenings that mostly do not have  \halpha\  wing counterparts. This indicates that the energy release witnessed by most of the EEs  occurs in the upper chromosphere and/or transition region of the solar atmosphere and it could be related to small-scale loop braiding at this height. A forward modeling with output of plasma observables  (e.g. spectral lines, images, etc.)  based on 3D MHD simulations of  magnetic reconnection in loop braidings is crucial to test this idea.  Although rare, chromospheric  events may eventually trigger explosive events in the transition region and in these cases they are characterized by absorption components (Fe\,{\sc ii} and Ni\,{\sc ii}) seeded in \siiv~1394\,\AA\ spectra. 

\section*{Acknowledgements}
 We thank the anonymous referee for his/her careful reading and constructive suggestions and comments.
This research is supported by the China 973 program 2012CB825601, and is supported by the National Natural Science Foundation of China (41404135, 41274178, 41474150), the Shandong provincial Natural Science Foundation (ZR2014DQ006) and the China Postdoctoral Science Foundation.  M.M. is supported by the Leverhulme Trust. Research at the Armagh Observatory is grant-aided by the N. Ireland Department of Communities, Arts and Leisure. IRIS is a NASA small explorer mission developed and operated by LMSAL with mission operations executed at NASA Ames Research center and major contributions to downlink communications funded by the Norwegian Space Center (NSC, Norway) through an ESA PRODEX contract. The Swedish 1-m Solar Telescope is operated on the island of La Palma by the Institute for Solar Physics of Stockholm University in the Spanish Observatorio del Roque de los Muchachos of the Instituto de Astrofsica de Canarias. The authors wish to acknowledge the DJEI/DES/SFI/HEA Irish center for High-End Computing (ICHEC) for the provision of computing facilities and support. We also like to thank the Solarnet project which is supported by the European Commission's FP7 Capacities Programme under Grant Agreement number 312495 for T\&S. The authors thank K. Galsgaard for the fruitful discussion on the manuscript. Z.H. and M.M. thank ISSI (bern) for supporting the team ``Solar UV bursts -- a new insight to magnetic reconnection''.

\bibliographystyle{mnras}
\bibliography{references1}

\appendix
\section{Explosive event line profile identification}
\label{sect_eeid}

EEs were named as "turbulent events" when they were firstly observed by HRTS and identified by the velocities of the wings of the transition region spectra ranged from 50$\sim$250\,\kms\,\citep{1983ApJ...272..329B}.  \citet{1989SoPh..123...41D} reported on the various shapes of C~{\sc iv} line profiles (see their Figures~2, 3, 6, 8, 9, 11, 13, and 15) that were named as ``explosive events''.  \citet{1989SoPh..123...41D} presented a statistical analysis of the Doppler shifts of EE spectra and showed a minimum shift of 50\,\kms. The Doppler shifts were mostly distributed within the 50--150\,\kms\ range. Variety of definitions have been used for the identification of EEs. For instance, \citet{1998ApJ...497L.109C} identified explosive events as spectra with non-thermal velocities of at least 45 km/s above average; \citet{2004A&A...427.1065T} have made a pre-selection by taking line profile parameters derived from single Gaussian fits exceeding 3 sigma; \citet{2008ApJ...687.1398M} identified EEs by calculating enhanced wings between 70--130\,\kms.

\par
In this study, explosive event line profiles are defined as the \siiv\ spectra with either enhanced wings or extremely broadened profiles (see definitions below). In order to extract them from the large number of spectra (105\,768 in total), an automatic detection technique has been developed.  Before describing the procedures, we  need to define the term  ``local peak'' of a line profile. A local peak is defined as a dispersion pixel in the observed line profile whose radiance is greater than the radiance in both or one of the neighboring pixels. If more than one neighboring pixel is identified as local peaks, they are considered as one. Only the local peaks with intensities larger than 10\% of the maximum of the line profile are considered. The algorithm follows five steps that are described below:

\par
(1) A pre-filter step: In the IRIS level 2 data, pixels with negative intensities can be found. This is caused by the background subtraction during the calibration process. These pixels are filtered out.
\par
(2) A noise filter step: Two other types of line profiles in the observations have to be filtered out before any further identification.  The first 
are line profiles that are polluted by high energy particles (e.g. cosmic rays, radiation belt particles, etc.). These spectra present as a sharp emission increase in one or more pixels and are usually called spikes. 
For a line profile $I(\lambda)$, we first identify its local peaks $I({\lambda_p})$, then we search for a minimum intensity $I_{min}(\lambda_p)$ within $\pm$3 dispersion pixels centered at $\lambda_p$, i.e. $I_{min}(\lambda_p)=min\{I(\lambda_p-3 : \lambda_p+3)\}$. If  the ratio $I(\lambda_p)/I_{min}(\lambda_p)$ for at least one local peak in the line profile is greater than 50, it is defined as a spike and the line profile is filtered out. 

\par
 Line profiles that  do not have a sufficient signal-to-noise ratio (SNR) are the second type of profiles to be removed. To identify these line profiles, we first determined the \siiv\ line center ($\lambda_{cr}$) from an average line profile ($I_{ref}$) of a quiet-Sun region. For each spatial pixel, we obtain the mean intensity of 5 dispersion pixels centered at  $\lambda_{cr}$, i.e. $I(\lambda_{cr})=mean\{I(\lambda_{cr}-2:\lambda_{cr}+2)\}$, which represents  the signal level of the spectral line. We then obtain the average intensity of a few dispersion pixels at  both the far blue  and red wings of the line profile, which represent the noise level, $I_{n}$. These wings correspond to Doppler velocity in the range from  194 to 218\,\kms, i.e. $I_{n}=mean\{I(\lambda_{cr+194}:\lambda_{cr+218}), I(\lambda_{cr-218}:\lambda_{cr-194})\}$. Spatial pixels for which the ratio $I(\lambda_{cr})/I_{n}$ is less than 50 are considered as low SNR  profiles and are excluded from further analysis.

\par
(3) Line width calculation step: All ($\sim$43\,000) spectra that have passed the pre-filter and noise filter steps are fitted with a single Gaussian function and their line widths ($w_\lambda$) are obtained. The average line profile derived from a quiet-Sun region, $I_{ref}$, is also fitted  with a Gaussian function and its line width is  denoted as $w_{\lambda,ref}$. The large range of spectra with various shapes due to intensity enhancements in the line wings that are defined as  EE spectra are given by  \citet{1989SoPh..123...41D}. Line profiles that have very strong wings with intensities comparable to the central component appear very broad rather than clearly wing-enhanced. Such line profiles have been reported in both HRTS C~{\sc iv} \citep[][their Figures\,13\&15]{1989SoPh..123...41D} and IRIS Si~{\sc iv} \citep[][their Figures\,S4--S6]{2014Sci...346C.315P} data. In our algorithm, these line profiles are also considered as non-Gaussian, and those whose width is greater than $3w_{\lambda,ref}$ are selected.

\par
The spectra that were not selected by the step (3)  are considered in the  next two steps that identify  spectra with  EE properties, i.e. enhanced line wings.

\par
(4) Enhanced wing extraction: Each spectrum, $I(\lambda)$  is  fitted with a single Gaussian function to obtain its line center ($\lambda_c$), line width ($w_\lambda$), and the background emission ($I_{bg}$). An artificial single Gaussian profile, $I'(\lambda$) is then constructed with a peak intensity $I'_p$, a line center $\lambda_c$, a line width $w_\lambda'$, and a background, $I_{bg}$, where 
the peak intensity, $I'_p$, is obtained as the $mean\{I(\lambda_c-2:\lambda_c+2)\}$; the line width, $w_{\lambda}'$ is selected as $w_\lambda$ in most of the cases, but is equated to $w_{\lambda,ref}$ when $w_\lambda< w_{\lambda,ref}$. Next, $I'(\lambda)$ is then subtracted from $I(\lambda)$ to obtain a difference (residual) spectrum, $I''(\lambda)$. If $I(\lambda)$ has enhanced wings, they will be identified as local peaks in $I''(\lambda)$. We denote the location of the local peaks in $I''(\lambda)$ as $\lambda_p''$. We use $w_{\lambda,ref}$ instead of $w_\lambda$ when $w_\lambda< w_{\lambda,ref}$ because a slightly small $w_{\lambda}'$ used in a very high intensity line with a Gaussian profile can easily lead to a large residual wing that can be identified as EE spectra (see next step).

\par
(5) EE spectra identification:  Here, EEs are identified as transition region emission spectra with enhancement in both the blue and the red wings at Doppler velocities more than 50\,\kms\ (for details, see Section\,\ref{sect_intro}).  Before the final identification, additional corrections need to be made to remove many small local (pixelwise) intensity peaks that are present  in the line profiles and are caused by various observational or instrumental effects. Therefore, we introduce two additional criteria in order to remove them. The first one is $I''(\lambda_p'')>0.1I_p'$, which eliminates low intensity local peaks. The second one is that the hump centred at $\lambda_p''$ has to be wider than $0.8w_{\lambda,ref}$, which eliminates narrow intensity spikes. The width of the spikes centred at $\lambda_p''$ is obtained from a single Gaussian fit of $I''(\lambda)|_{\lambda=\lambda_p''-68\rightarrow\lambda_p''+68}$, which is a portion of $I''(\lambda)$ covering $\pm$68\,\kms\ Doppler shifts on both sides of $\lambda_p''$. The Doppler shift of 68\,\kms\ is twice the full width at half maximum (FWHM) of the reference spectrum, which is about 34\,\kms. If the above three criteria are met, the spectrum $I(\lambda)$ is selected as an EE spectrum.

\bsp	
\label{lastpage}
\end{document}